# Against normality testing

Paolo Frumento
University of Pisa
Department of Political Sciences
Via Serafini 3, 56126 Pisa, Italy
paolo.frumento@unipi.it

> *I can only recognise the occurrence of the normal curve [...] as a very abnormal phenomenon. It is roughly approximated to in certain distributions; for this reason, and on account for its beautiful simplicity, we may, perhaps, use it as a first approximation, particularly in theoretical investigations.*
>
> **Karl Pearson, 1901**

**Abstract:** I reject the following null hypothesis: {$H_0$: your data are normal}. Such drastic decision is motivated by theoretical reasons, and applies to your current data, the past ones, and the future ones. While this situation may appear embarrassing, it does not invalidate any of your results. Moreover, it allows to save time and energy that are currently spent in vain by performing the following unnecessary tasks: (i) carrying out normality tests; (ii) pretending to do something if normality is rejected; and (iii) arguing about normality with Referee #2.

**Introduction**

The normal distribution (also known as Gaussian distribution, Laplace's second law, or law of error) was first introduced by J.C.F. Gauss in his 1823 monograph *Theoria combinationis observationum erroribus minimis obnoxiae.* In the context of Gauss' work, *normal* had the technical meaning of "orthogonal"; soon, however, it was interpreted as "common" or "typical". Quetelet (1835) introduced the idea that human traits, such as weight or intelligence, are well described by mean and variance and follow a normal distribution. Similar concepts were presented by Galton (1877) in his work on population stability.

These early observations showed that various real-world quantities have a unimodal, bell-shaped, vaguely symmetrical distribution. However, there is no theoretical reason to believe that normality is the norm, and nature and literature are rich with asymmetric distributions and fat tails. The existence of a misconception about normality was already recognised by Pearson in his 1901 and 1920 articles:

*Many years ago I called the Laplace-Gaussian curve the normal curve, which name, while it avoids an international question of priority, has the disadvantage of leading people to believe that all other distributions of frequency are in one sense or another 'abnormal' (Karl Pearson, 1920).*

In reality, not only the normal distribution is not the norm; it does not describe *exactly* real world phenomena, if not in particular cases.[1] Therefore, the null hypothesis

$$H_0: \text{your data are normal}$$

can be safely assumed to be false and rejected in advance, regardless of what the data suggest.

The above claim has important consequences on the way we treat our data. From a strictly theoretical standpoint, the common practice of two-stage testing, i.e., the procedure of "verifying" normality before carrying out t-test or ANOVA, appears nonsensical. If we forgive this logical aspect and focus on sheer performance, two-stage testing will often lead to unnecessary or even detrimental actions. In particular, normality will always be rejected when the sample size is sufficiently large, i.e., precisely when we care less about normality; instead, when the sample size is very small and the validity of assumptions may actually affect performance, normality may not be rejected due to insufficient power.

The good news is that parametric tests work well even with non-normal data, and the concerns expressed by Referee #2 on your choice of using t-test are unjustified. Alternatively, you can apply a nonparametric test and avoid this type of discussions[2]. If you are worried to lose power by replacing t-test with Wilcoxon rank-sum test, be aware that this concern is also unjustified. Contrary to common belief, nonparametric methods have only slightly less power than parametric approaches, when the data are sufficiently symmetric or the sample size is extremely small, but are often much more powerful when the distribution is skewed and the sample size is larger than that of your fingers.

This paper is structured as follows. In Section 1, I justify my claim of universal non-normality. In Section 2, I criticise the current practice regarding two-stage testing, present simulation results, and provide some general guidelines.

---

[1] Exact normality can be found in particle and quantum physics, where a precise theory describes all the involved masses, motions, forces, charges, etc.

[2] Actually, the best approach is to use a parametric test in the first submitted version of your manuscript, which will lure Referee #2 into criticising your choice and distract him/her from other, more relevant flaws. Then, in the revised version, you convert to nonparametric tests (following the Referee's comment…).

**1. Normality is not normal**

The normal distribution plays a fundamental role in statistical theory. By virtue of central limit theorem, it describes the asymptotic behaviour of estimators such as the sample mean, ordinary least squares, and maximum likelihood estimators; additionally, it can be used to conveniently describe random errors, making a somewhat plausible assumption to simplify problems that would otherwise be analytically intractable.

However, the normal distribution is *not* meant to describe a particular random variable X, such as the IQ or the body mass index. While some real-world distributions *appear* normal, this is not a "proof" of normality. Using Curran-Everett's (2017) words: "*Although the data themselves can be consistent with a normal distribution, they need not be*".

The implicit notion that "X is normal, unless proven otherwise", which is also the reasoning behind testing normality, constitutes a textbook example of "reverse burden of proof". The reality is that X is *not* normal, unless you can prove it. Clearly, *not rejecting a normality test does not prove normality*. The normal model has a precise mathematical definition as a limiting distribution; proving that X is normal would require to show that the underlying data-generating process *tends* to a certain functional. Such a proof is unlikely to exist, as it would entail some form of convergence towards a limit, a concept that can hardly be applied to real-world quantities.

The conclusion is that, in absence of a strong theoretical motivation, there is *no reason* to believe that X has exactly this distribution:

$f_X(x) = \frac{1}{\sigma\sqrt{2\pi}} e^{-\frac{(x-\mu)^2}{2\sigma^2}}$, a normal distribution,

and not, for example,

$f_X(x) = \frac{1}{\Gamma(\alpha)\beta^\alpha} x^{\alpha-1} e^{-x/\beta}$, a gamma distribution,

or maybe a logistic, beta, exponential, or Student's t distribution, *or, most likely, none of the above*[3].

The main claim of this paper is that the null hypothesis {$H_0$: X is normal} is not true, unless there is a valid reason to believe otherwise. Usually, no such reason exists for biological, economic, or socio-demographic variables.

Obviously, researchers should not test a hypothesis that is known in advance to be false. This is summarised in the following Rhetorical Question:

> **RQ1: Why test normality, if normality is known to be false?**

One may object that applying a normality test, although nonsensical, can do little harm and may actually improve inference, for example by suggesting to use a nonparametric test instead of a parametric one. However, as shown in Section 2, the test results can be quite misleading and should not be used to guide decisions regarding data analysis.

---

[3] As a side note, there are numerous families of random variables that, for suitable values of the parameters, generate a bell-shaped, symmetric distribution: for example the gamma, beta, and Weibull. None of these is characterised by location and scale parameters, that play a fundamental role in the normal distribution; the whole idea of "mean + deviation" might not be how nature generates data.

## 2. Two-stage testing

Before performing t-test or ANOVA, researchers "verify"[4] that, within each group, the X variable is normally distributed. If normality is rejected, they may transform the data or use a nonparametric test (e.g., Rochon, Gondan, and Kieser 2012; Shamsudheen and Hennig 2019), such as Wilcoxon rank-sum or Kruskal-Wallis test[5]. This combined approach is often referred to as "two-stage testing".

Most scholars recognise that exact normality is totally unrealistic. However, they generally agree with the idea of two-stage testing, and interpret the preliminary normality test as a method to quantify deviations from normality. Quoting Shamsudheen and Hennig (2019):

"… *model assumption checking needs to distinguish problematic violations from unproblematic ones, rather than distinguishing a true model from any wrong one* […] *It is as misleading to claim that model assumptions are required to hold (which is an ultimately impossible demand) as it is to ignore them, or rather to ignore potential performance breakdown of the procedure to be applied on models other than the assumed one*".

However, the p-value of a normality test is *not* the correct instrument to "distinguish problematic violations from unproblematic ones". The performance breakdown of a test that relies on normality is an increasing function of the distance between the data and the gaussian distribution, and a decreasing function of the sample size; *instead, the p-value of a normality test is a decreasing function of both*.

The implication is that, besides the theoretical arguments proposed in Section 1, there are strong practical reasons to dismiss two-stage testing. In Section 2.1 I show that two-stage testing leads to "take actions" to counteract non-normality precisely when those actions are the least necessary. In Section 2.2 I present a small simulation study, and show that preliminary normality testing is not necessarily harmful, but does not constitute a valid decision rule.

### 2.1 The paradox

Because X is never exactly normal, the result of a normality test depends essentially on the available sample size[6]. Intuitively, normality *will always be rejected* as long as *n* is large enough; and it might not be rejected, causing researchers to incur in type-II error, whenever *n* is too small to achieve a sufficient statistical power.

Consider the two simulated datasets illustrated in Figure 1. Albert (on the left) has a sample of $n = 25$ observations. His distribution appears asymmetric and clearly not normal; this, combined with the small sample size, may yield incorrect inference. On the other hand, Bernie (on the right) has a sample of $n = 1000$ data points. Not only he has a much larger sample size, but also a bell-shaped, symmetric distribution; an ideal situation in which standard asymptotic theory will produce reliable results.

Unfortunately, if we follow the standard practice and use Shapiro-Wilk test to guide our decisions, we reach the opposite conclusions. In Albert's case, due to lack of power, we do not reject normality at a significance level α = 0.05 (or α = 0.10, for that matter). With Bernie's data, instead, normality is rejected with a p-value of 0.002. This may lead Referee #2 to not question Albert's results, while asking Bernie to "do something" to handle non-normality.

---

[4] This statement is formally inappropriate and, unfortunately, rather common. Not only it conveys the wrong idea that data should be expected to be normal, unless evidence of the contrary is found; it also suggests that a null hypothesis can be "proved", which is not how statistics (and epistemology) works.

[5] It should be clarified that parametric tests and rank-based methods, that are often seen as "substitutes", do not test the same null hypothesis. See Fay and Proschan (2010) for a discussion. If you are *really* into comparing *means*, you can stop reading this and just do a t-test.

[6] Clearly, for any given sample size, the power of the test will also depend on the actual data distribution.

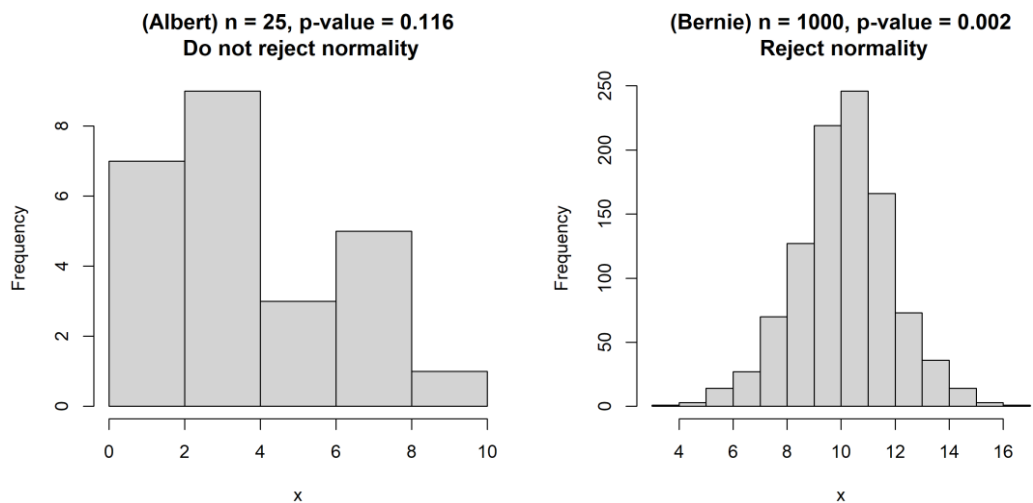

**Figure 1. Two simulated datasets. Albert has *n* = 25 observations and does not reject normality (p = 0.116). Bernie has *n* = 1000 observations and rejects normality (p = 0.002).
In reality, both datasets are generated from non-normal distributions.**

The reason for the paradox is that even a small deviation from normality could be highly significant, if *n* is large; and vice versa, a large deviation might not be detected, if *n* is small. Bernie might feel compelled to take unnecessary actions, such as transforming the data or applying a nonparametric test. In Albert's scenario, where these actions might actually improve inference, the common practice is to claim normality (which is, *per se*, a wrong interpretation of the test results) and apply a t-test to the untransformed data.

Similar arguments are presented, among others, by le Cessie, Goeman and Dekkers (2020). Using the words by Lumley et al (2002): "*Formal statistical tests for Normality are especially undesirable as they will have low power in the small samples where the distribution matters and high power only in large samples where the distribution is unimportant.*"

To exemplify the problem with real data, I conducted a small experiment. I considered all numerical variables available in the `datasets` R package, version 4.3.0. The package includes various datasets such as `iris`, `cars`, and `faithful`. I removed count variables that were "too small" to be treated as continuous, and data with repeated measures; I replaced time series by their first differences; when possible, I considered both the original variable and its logarithm. Finally, I stratified all datasets by the relevant factors (e.g., treatments or exposure groups). In total, I obtained 292 variables to which I applied Shapiro-Wilk test for normality. The R code is provided as supplementary material.

As shown in Figure 2, there is a clear relationship between sample size (on the x-axis) and the p-value of the test (on the y-axis). Rejecting normality is rare when *n* is small, and becomes the most frequent outcome as *n* increases. In our example, the proportion of cases with p-value less than 0.05 was 24%, when $n \leq 30$, and 72%, when $n > 50$. None of the 14 variables with sample size greater than 250 passed the normality test. Clearly, this does not mean that larger samples are less likely to be normal. It only shows that, as *n* increases, rejecting normality becomes inevitable. The reason behind this is that, in reality, no variable is normally distributed.

The above experiment provides (unnecessary) evidence for the claim of universal non-normality presented in Section 1. Using the same intuition, Albert and Bernie must admit that their samples are not generated from a normal distribution, whether the data agree or not. This realisation, however, does not tell us what to do. To provide some guidelines, in the next section I present a simulation study. At the end, we still do not know what to do, but in a more scientific way.

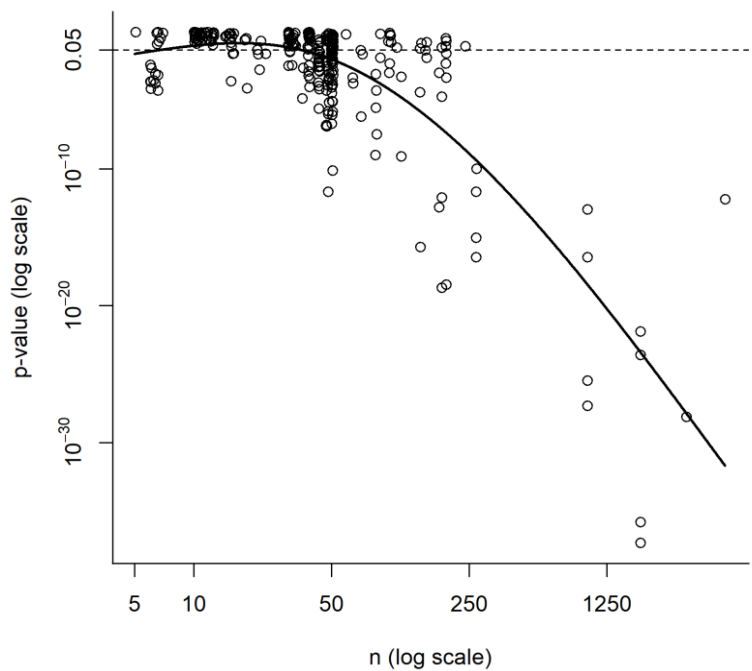

**Figure 2. Relationship between sample size (x-axis) and p-value of Shapiro-Wilk test for normality (y-axis) using the numerical variables available in the `dataset` R package.**
**The horizontal line corresponds to p = 0.05.**

**2.2 Simulation results**

In Section 1 we suggested that, in reality, data are never generated by a normal distribution. This can be used as an argument to claim that normality testing is pointless: the null hypothesis is always false, and not rejecting it is not a "proof" of normality, by any stretch of imagination. If the goal is to identify which deviations from normality should be regarded as "problematic", in Section 2.1 we showed that the chance of rejecting normality is inversely proportional to how problematic non-normality is. This ultimately results in bad decision making when it comes to two-stage testing procedures.

However, from the practitioners' point of view, two-stage testing might be better, in terms of type-I or type-II error rates, than blindly choosing which test to apply. As they say, if it works, why not. This explains why positions against two-stage testing (e.g., Fay and Proschan 2010; Rochon, Gondan, and Kieser 2012) are generally based on performance, and not on *a priori* rejection of the null hypothesis.

In this section, we present a small simulation study to illustrate what kind of performance we can expect from a parametric test, a nonparametric test, and a two-stage test procedure. While the simulation is far from being exhaustive, it may help to formulate guidelines. More extensive research in this direction is already presented in the literature, e.g., Boneau (1960), Lumley et al (2002), Huang and Qu (2006), Rasch, Teuscher and Guiard (2007), Shuster (2009), Fay and Proschan (2010), Rochon, Gondan and Kieser (2012), Curran-Everett (2017), Shamsudheen and Hennig (2019), le Cessie, Goeman and Dekkers (2020), Knief and Forstmeier (2021), among others.

The simulation evaluates the performance of two-sample tests under different data distributions: normal, logistic, uniform, gamma (shape = 2, scale = 1), exponential, and a normal mixture, 0.9·N(0, 1) + 0.1·N(7.5, 1), that generates a small cluster of outliers. Under the null, the two samples are simulated from the same distribution. Under the alternative hypothesis, the two distributions differ by a location shift equal to 0.5 standard deviations. Three different testing approaches are considered: t-test with assumed equal variance (T-test); Wilcoxon rank-sum test (W-test); and the combined procedure (C-test), in which we apply Wilcoxon test, if normality is rejected with $p < 0.05$, and T-test otherwise. Normality is assessed by Shapiro-Wilk test applied to the pooled residuals from the sample means. The R code is provided as supplementary material.

For different values of *n*, that denotes the size of *each* sample, Table 1 reports the chance of rejecting normality, i.e., the power of the Shapiro-Wilk test, and the performance of the two-sample tests, measured by the risk of type-I error, under the null hypothesis, and by the statistical power, under the alternative (nominal significance level: $\alpha = 0.05$).

Results can be summarised as follows.

- The risk of type-I error is always very close to its nominal value of 0.05, irrespective of the data distribution. With a moderate sample size (e.g., 25 observations in each group), non-normality and even severe skewness do not hinder the performance of T-tests. An error rate of 0.03-0.04, less than the nominal value, is only found under the following circumstances: in W-test, when *n* is very small (e.g., five or ten observations in each group); and in T-test, when *n* is very small *and* the data distribution is asymmetric. The fact that the risk of type-I error is *less* than expected can hardly be thought of as a problem; however, one may argue that parametric tests should be preferred with very small samples.

- With only 5 observations per group, T-test has more power than W-test, except in the normal mixture where the distribution is very, very asymmetric. With an acceptable sample size, T-test is only slightly preferable with light-tailed distributions (normal and uniform), while W-test is more powerful, often by a large margin, in all other situations.

- The combined procedure gives mixed signals. In terms of type-I error, C-test grants some minor improvement over T-test and W-test only when the distribution is very asymmetric (exponential and normal mixture) and the sample size is extremely small. In terms of power, results go in a variety of directions. With normal data, C-test is not better than T-test. With symmetric, light-tailed data (uniform distribution), T-test is preferable; in this scenario, preliminary normal testing often suggests applying W-test, which is not the optimal decision. No clear winner is found in the logistic scenario. In all other situations, where the distribution is asymmetric, the combined procedure suggests W-test with a high probability, with the result that W-test and C-test are essentially equivalent.

There are three important considerations to be made: (i) in terms of type-I error, concerns about the validity of parametric tests with non-normal data are largely unjustified; (ii) in terms of power, nonparametric tests can be *slightly* worse, but also *much* better than parametric approaches; and (iii) the combined procedure based on preliminary normality testing might or might not be preferable, depending on the situation; however, simulation evidence points towards a different decision rule in which the critical ingredients are sample size, asymmetry, and tail behaviour ("size" of the tail and presence of outliers).

Regarding (i), parametric tests are well known to be robust to deviations from normality (e.g., Boneau 1960, Rasch, Teuscher and Guiard 2007, and, in a more general class of models, Knief and Forstmeier 2021). In our simulation, T-test has a correct type-I error rate, even with asymmetric data, with some performance breakdown only when *n* is very small[7].

Regarding (ii), when W-test is *less* powerful than T-test, the difference does not exceed 2 to 4 percentage points. Vice versa, when W-test is *more* powerful than T-test, the gain is likely to be substantial. For example, with normal data and $n = 25$, the power of W-test is 0.39, and that of T-test is 0.41; with a normal mixture, W-test has a power of 0.90, while T-test lags behind with 0.43. These results are consistent with the existing literature (e.g., Huang and Qu 2006). With this in mind, always applying a nonparametric test appears to be a meaningful option: in the worst case, we may suffer from a minor loss; and in the best case, we can enjoy a large gain.

However, point (iii) above suggests that there is a better approach, in which the choice between a parametric test and a nonparametric one is guided by criteria that have little to do with the outcome of a normality test. As shown by the simulations, there are only two situations in which T-test outperforms W-test: either when *n* is really small (e.g., le Cessie, Goeman and Dekkers 2020); or when the distribution is symmetric and short-tailed (normal, uniform, or similar). In these situations, T-test shows a better power and, with small *n*, a type-I error rate closer to the nominal level. The suggested decision rule is: *always use a nonparametric test, unless one of the following conditions is met: either n is extremely small (e.g., less than 10), or the data are sufficiently symmetric, have short tails, and do not include outliers.*

Someone may find this recommendation too vague. Often, practitioners want a precise rule: for example, something like "use a parametric test only if the absolute value of the sample skewness is less than $4.8/\sqrt{n}$, or if $n \leq 12$". However, such rules are often arbitrary[8] and cannot be valid in general. According to the Internet, the best approach is to choose which test to use based on a visual inspection of the relevant histograms (Shamsudheen and Hennig 2019). This does not sound rigorous, but might still be a better idea than using a fixed rule. If, like me, you believe that this is not taking us anywhere, just do as follows: (a) try to have more than just 5 observations per group; and (b) provided that (a) is true, always use nonparametric tests, or follow footnote 2.

But please, stop testing normality.

---

[7] Based on my personal sensibility, knowing that the risk of type-I error could be 3.5% instead of 5% is completely irrelevant. One reason, that has already been mentioned, is that 3.5% is smaller, and not larger, than 5%. The main reason, however, is that 5% does not mean anything except "a small, round number in base 10". Pandas, who have 6 fingers, count in base 12 and like $1/24 = 0.041\bar{6}$ more than $1/20 = 0.05$. While this may appear surprising, it has nothing to do with their impending extinction.

[8] or even made up, as in this case.

|  |  |  | Type I error | | | Statistical power | | |
| --- | --- | --- | --- | --- | --- | --- | --- | --- |
| Distribution | n | rej norm | T-test | W-test | C-test | T-test | W-test | C-test |
| Normal | 5 | 0.04 | 0.05 | 0.03 | 0.05 | 0.11 | 0.07 | 0.10 |
| 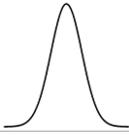 | 10 | 0.05 | 0.05 | 0.04 | 0.05 | 0.18 | 0.16 | 0.19 |
|  | 25 | 0.05 | 0.05 | 0.05 | 0.05 | 0.41 | 0.39 | 0.41 |
|  | 50 | 0.05 | 0.05 | 0.05 | 0.05 | 0.70 | 0.68 | 0.70 |
|  | 100 | 0.05 | 0.05 | 0.05 | 0.05 | 0.94 | 0.93 | 0.94 |
| Logistic | 5 | 0.06 | 0.05 | 0.03 | 0.05 | 0.11 | 0.08 | 0.11 |
| 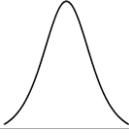 | 10 | 0.10 | 0.05 | 0.04 | 0.05 | 0.19 | 0.18 | 0.20 |
|  | 25 | 0.19 | 0.05 | 0.05 | 0.05 | 0.42 | 0.44 | 0.44 |
|  | 50 | 0.30 | 0.05 | 0.05 | 0.05 | 0.70 | 0.73 | 0.72 |
|  | 100 | 0.49 | 0.05 | 0.05 | 0.05 | 0.94 | 0.95 | 0.95 |
| Uniform | 5 | 0.06 | 0.05 | 0.03 | 0.05 | 0.10 | 0.06 | 0.10 |
| 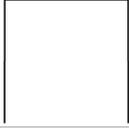 | 10 | 0.13 | 0.05 | 0.04 | 0.05 | 0.17 | 0.15 | 0.18 |
|  | 25 | 0.62 | 0.05 | 0.05 | 0.05 | 0.40 | 0.37 | 0.39 |
|  | 50 | 0.98 | 0.05 | 0.05 | 0.05 | 0.69 | 0.65 | 0.65 |
|  | 100 | 1.00 | 0.05 | 0.05 | 0.05 | 0.94 | 0.91 | 0.91 |
| Gamma | 5 | 0.15 | 0.04 | 0.03 | 0.04 | 0.12 | 0.09 | 0.12 |
| 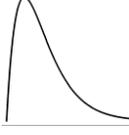 | 10 | 0.43 | 0.05 | 0.04 | 0.05 | 0.20 | 0.22 | 0.24 |
|  | 25 | 0.92 | 0.05 | 0.05 | 0.05 | 0.43 | 0.54 | 0.54 |
|  | 50 | 1.00 | 0.05 | 0.05 | 0.05 | 0.70 | 0.83 | 0.83 |
|  | 100 | 1.00 | 0.05 | 0.05 | 0.05 | 0.94 | 0.99 | 0.99 |
| Exponential | 5 | 0.27 | 0.04 | 0.03 | 0.04 | 0.14 | 0.12 | 0.14 |
| 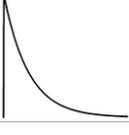 | 10 | 0.70 | 0.04 | 0.04 | 0.05 | 0.23 | 0.30 | 0.31 |
|  | 25 | 1.00 | 0.05 | 0.05 | 0.05 | 0.44 | 0.69 | 0.69 |
|  | 50 | 1.00 | 0.05 | 0.05 | 0.05 | 0.71 | 0.94 | 0.94 |
|  | 100 | 1.00 | 0.05 | 0.05 | 0.05 | 0.94 | 1.00 | 1.00 |
| Mixture | 5 | 0.47 | 0.03 | 0.03 | 0.04 | 0.18 | 0.21 | 0.23 |
| 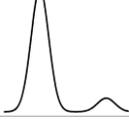 | 10 | 0.86 | 0.04 | 0.04 | 0.05 | 0.27 | 0.48 | 0.49 |
|  | 25 | 0.99 | 0.05 | 0.05 | 0.05 | 0.43 | 0.90 | 0.90 |
|  | 50 | 1.00 | 0.05 | 0.05 | 0.05 | 0.70 | 1.00 | 1.00 |
|  | 100 | 1.00 | 0.05 | 0.05 | 0.05 | 0.94 | 1.00 | 1.00 |

**Table 1. Performance of two-sample tests under different data distributions (see text for details). The value of *n* represents the size of *each* sample. T-test: t-test assuming equal variance; W-test: Wilcoxon rank-sum test; C-test: combined procedure. The table reports the probability of rejecting normality; the risk of type-I error, and the statistical power when the true location shift is equal to 0.5 standard deviations.**

**Final remarks**

The main goal of this article was to convince as many people as possible that testing normality is nonsensical. There is no way that your data are normal, and the null hypothesis can be rejected in advance.

If this was not a sufficient argument, I showed that normality testing is only effective when it is unnecessary. The validity of parametric tests depends on a combination of symmetry, tail behaviour, presence of outliers, and sample size[9]. The p-value of a normality test is a hotchpotch of these ingredients; in particular, it is a decreasing function of both the sample size and the distance between the data and the normal distribution, two factors that have opposite effects on the performance of parametric procedures. As a result, you will most likely reject normality when *n* is large and normality is no longer an issue; and vice versa, you may not reject normality, due to insufficient power, precisely when a nonparametric test could be preferable.

Finally, the fact that standard two-stage testing has an acceptable performance does not make it the correct approach. I tried to suggest some alternative decision rules, but I could only come up with vague guidelines. In absence of a simple rule that combines all the relevant ingredients, I recommend to always use nonparametric tests. This might not be a good idea if you only have 3 observations per group; however, with such small sample size, you could as well give up your research.

In any case, I really hope you will stop testing normality.

---

[9] as well as the plausibility of homo/heteroskedasticity assumptions, that were never mentioned but are much more important than normality.